\documentclass[12pt]{iopart}

\usepackage{graphicx}
\usepackage{iopams}  
\usepackage{xcolor}

\begin{document}


\title[]{Energy and system size dependence of strongly intensive fluctuation measures in heavy-ion collisions at FAIR energies}
\author{Bushra Ali$^{1}$, Shakeel Ahmad$^{1}$ and A. Ahmad$^{2}$}

\address{$^{1}$Department of Physics, Aligarh Muslim University, Aligarh-India \\
		 $^{2}$Department of Applied Physics, Aligarh Muslim University, Aligarh-India}
		 
\ead{bushra.ali@cern.ch, shakeel.ahmad@cern.ch}
\vspace{10pt}

\begin{indented}
\item[] Accepted Manuscript online 10 November 2023\\

\item[] {\footnotesize $\copyright$}  2023 IOP Publishing Ltd
\item[]{This Accepted Manuscript is available for reuse under a CC BY-NC-ND licence after the 12 month embargo period provided that all the terms and conditions of the licence are adhered to}

\end{indented}

\begin{abstract}
Event-by-event fluctuations of multiplicity and transverse momentum of charged hadrons produced in heavy-ion collisions at {\footnotesize FAIR} energies, 10A, 20A, 30A and 40A GeV are studied in the framework of relativistic transport model, {\footnotesize URQMD}. Dependence of two families of strongly intensive measures of multiplicity($N$) and transverse momentum($p_{\rm T}$) fluctuations, $\Delta[p_{\rm T},N]$ and $\Sigma[p_{\rm T},N]$, on collision centrality, centrality bin-widths and pseudorapidity windows are examined. Attempts are also made to study $NN$, $N$$p_{\rm T}$ and $p_{\rm T}$$p_{\rm T}$ fluctuations using two window analysis method. The findings suggest that the measure, $\Delta[p_{\rm T},N]$ be dealt with proper selection of centrality intervals. This measure also exhibits a strong dependence on the widths of $\eta$ windows. The variable $\Sigma[p_{\rm T},N]$, however, is observed to be insensitive to the centrality bin-widths and shows a variation of $< 5\%$ with the widths of $\eta$ windows. The analysis of data after event mixing gives $\Delta[p_{\rm T},N]$ and $\Sigma[p_{\rm T},N]$ values as $\sim 1$ irrespective of the widths of $\eta$ windows and collision centrality, as predicted by model of independent particle emission, IPM. The study of joint fluctuations of the two quantities on two $\eta$ windows separated in $\eta$ space, reveals that $\Sigma[N_{\rm F},N_{\rm B}]$ values are $\sim 1$ irrespective of the position of $\eta$ windows whereas, the values of $\Sigma[N_{\rm F},p_{\rm T_B}]$ and $\Sigma[p_{\rm T_F},p_{\rm T_B}]$ firstly increase with $\eta_{sep}$ and later acquire saturations. The observed trend of centrality dependence of $\Sigma[N_{\rm F},N_{\rm B}], \Sigma[N_{\rm F},p_{\rm T_B}]$ and $\Sigma[p_{\rm T_F},p_{\rm T_B}]$ agrees fairly well with those observed in MC simulated studies carried out for AA collisions at LHC energies in the framework model of string fusion.
\end{abstract}

\vspace{2pc}

\newpage

\section{Introduction}
One of the challenging tasks in the studies of relativistic and ultra-relativistic heavy-ion({\footnotesize AA}) collisions is to explore the phase diagram of strongly interacting matter by exploring the possible phase boundaries and critical point({\footnotesize CP})\cite{label1,label2}. In the vicinity of {\footnotesize CP}, fluctuations in particle multiplicities are envisaged to be enhanced which provide means of correlating experimental measures to the critical behavior of the system created in {\footnotesize AA} collisions\cite{label3}. However, in {\footnotesize AA} collisions, the volume of the produced matter is not a fixed quantity as the impact parameter varies from event-to-event, which produces a statistical system having the same local properties, like temperature, {\footnotesize T} and baryonic chemical potential $\mu_B$ but the volume of the system would vary from event-to-event\cite{label4}. It is, therefore, preferred to address the multiplicity fluctuation studies in terms of intensive and strongly intensive variables. The intensive variables do not depend on system volume, whereas the strongly intensive variables are independent of volume fluctuations. The two families of strongly intensive variables, $\Sigma$ and $\Delta$ and the intensive variable, $\omega$(scaled variance) have been frequently employed in recent years to study the multiplicity fluctuations.\\

\noindent Several attempts have been made to address the multiplicity and transverse momentum fluctuations in hadronic and heavy-ion collisions at {\footnotesize SPS}, {\footnotesize RHIC} and {\footnotesize LHC} energies using these variables. The main objective of these investigations is to scan the two dimensional(T-$\mu_B$) phase diagram\cite{label2, label3, label5, label6, label7, label8, label9, label10, label11}. It has been reported[5] that non-trivial effects evolve from the Poissonian like fluctuations for narrow pseudorapidity($\eta$) intervals with expansion of the acceptance. These fluctuations are regarded to be sensitive to the existence of {\footnotesize CP. NA49} findings on inclusive hadron production indicate that the onset of deconfinement in central Pb-Pb collisions is located at $\sqrt{s_{NN}} \sim$ 30A GeV\cite{label2}. It is mainly based on the observation of narrow structures in the energy dependent hadron production in central Pb-Pb collisions\cite{label12, label13}. {\footnotesize STAR} collaboration in its beam energy scan of several observables, including ebe net proton distribution has reported\cite{label14, label15} an increase of $4^{th}$ order cumulants(kurtosis) towards lowest collision energy, $\sqrt{s_{NN}}$ = 7.7 GeV, which is, incidentally the lowest available energy in the collider mode. It has also been suggested\cite{label14, label15} that to explore the maximum of fluctuations and hence the possible location of {\footnotesize CP}, investigations ought to be extended to further lower energies. Such energy regions are expected to be available in future heavy ion experiments, {\footnotesize FAIR}(Facility of Antiproton and Ion Research). It may be mentioned here that the onset of deconfinement are visualized to occur in the collisions of heavier nuclei, like Pb, with beam energy, $E_{Lab} \sim$ 30A GeV\cite{label16, label17}.  \\

\noindent According to lattice {\footnotesize QCD} calculations\cite{label18, label19} at vanishing baryon-potential, $\mu_B$, the required critical temperature $T_{c}$ for {\footnotesize QGP} formation should be $\sim$ 170 MeV, which corresponds to an energy density of a few GeV/$fm^3$. The intermediate fireball produced at {\footnotesize FAIR} energies are expected to be in the higher regions of {\footnotesize QCD} phase diagram. Therefore, if a {\footnotesize QGP} hadron phase transition occurs, the transition temperature would be significantly lower and $\mu_B$ will be higher than those required for hadronization to take place from a baryon-free {\footnotesize QGP}\cite{label17}. The {\footnotesize CBM}(Compress Baryonic Mass) experiment at {\footnotesize FAIR} is expected to provide an opportunity to explore the high $\mu_B$ region of the phase diagram and to locate the {\footnotesize CP}\cite{label17, label18}. Non monotonic behavior of various fluctuation measures is regarded as a signal of {\footnotesize CP} as the magnitude of fluctuations and correlations may change significantly during phase transition, particularly for {\footnotesize AA} collisions which are passing close to {\footnotesize CP}\cite{label14, label20}. It is, therefore considered worthwhile to undertake a systematic study of event-by-event(ebe) multiplicity fluctuations in Au-Au collisions at future {\footnotesize FAIR} energies. Since no experimental data are available at these energies, data in the frame work of relativistic transport model are generated and analyzed to examine the dependence of strongly intensive quantities on the collision centrality and pseudorapidity windows of varying widths and separations. The {\footnotesize MC} data used, corresponds to {\footnotesize Au-Au} collisions at beam energies 10A, 20A, 30A and 40A GeV and are simulated using the code {\footnotesize URQMD} 3.4\cite{label21}.\\

\noindent Since ebe fluctuation studies carried out in the frame work of relativistic transport models like {\footnotesize EPOS}\cite{label22, label23}, {\footnotesize PHSD}\cite{label24, label25}, {\footnotesize URQMD}\cite{label21}, {\footnotesize AMPT}\cite{label26} and the models with quark-gluon string extended in $\eta$ space and acting as particle emitting sources\cite{label11} have been observed to describe nicely the available data from {\footnotesize SPS}, {\footnotesize RHIC} and {\footnotesize LHC}. The present study could provide a mean to examine the dependence of ebe fluctuation measures on collision centrality and phase space width, that might improve our understanding about the ebe fluctuations in baryon-rich environment and it might lead to some physics insight from the perspective of {\footnotesize CBM} experiment. \\

\section{URQMD Model}

{\footnotesize URQMD} model describes the phenomenology of hadronic interactions at low and intermediate energies ($\sqrt{s} <$ 5 GeV) in terms of interactions between hadrons and resonances. At high energies ($\sqrt{s} >$ 5 GeV), the excitation of color strings and their subsequent fragmentation into hadrons dominate the multiple productions of particles\cite{label21}. In a transport model, {\footnotesize AA} collisions are considered as the superposition of all possible binary nucleon-nucleon(nn) collisions and each nn collision with impact parameter, b $ < \sqrt{\sigma_{tot}/\pi}$, is taken into account, where $\sigma_{tot}$ denotes the total cross section. Two colliding nuclei are described by Fermi gas model and hence the initial  momentum of each nucleon is taken at random between zero and Thomas Fermi momentum\cite{label17, label27}.  The interaction term includes more than 50 baryon and 45 meson species\cite{label3}. The model preserves the conservation of strangeness, baryon number and electric charge\cite{label28}. The space-time evolution of the fireball is studied in terms of excitation and fragmentation of color strings and the formation and decay of hadronic resonances are accounted\cite{label28, label29}. The model also includes multiple scattering between hadrons during evolution, including baryon stopping phenomenon, which is an important feature of {\footnotesize AA} collision, particularly at lower energies\cite{label21}. The model has been successfully applied to study the thermalization\cite{label30}, particle yields\cite{label31, label32}, leptonic and photonic probes\cite{label33} and ebe fluctuations\cite{label34, label35, label36, label37, label38, label39}. Although the phase transition from hadronic to partonic phases are not explicitly included in the model, yet the study is expected to help in the interpretation of the experimental data as it would allow subtraction of simple dynamical and geometrical effects from the expected {\footnotesize QGP} signals\cite{label27, label40}. \\

\begin{table}
\begin{center}
\lineup
\item[]\begin{tabular}{@{}*{2}{l}}
\br                              
$\0\0E_{Lab}$&$Events$\cr 
$\0\0(GeV)$&$(10^6)$\cr 
\mr
\0\010A &	\08.3 \cr
\0\020A &	\08.0\cr 
\0\030A &	\06.8\cr 
\0\040A & 	\07.4\cr 
\br
\end{tabular}
\caption{\label{tabone} Number of events analyzed at various beam energies.} 
\end{center}
\end{table}

\section{Definition of observables}

A quantity which is proportional to the system volume is referred to as the extensive variable, whereas the one which does not depend on the system volume is termed as the intensive variable. The ratio of two extensive variables will result into an intensive variable, for example, mean charged particle multiplicity $\langle N\rangle$, is an extensive observable while the ratio of mean multiplicities of two different types of particles would lead to an intensive observable. Similarly, particle number fluctuations, which are quantified by the variance, $V (= \langle N^{2} \rangle - \langle N \rangle^2)$ is an extensive quantity, whereas, the scaled variance, $\omega (= V/\langle N \rangle)$ is an intensive one\cite{label4}. The volume of created matter in high energy collisions can not be kept fixed, i.e., in {\footnotesize AA} collisions with different centrality the events produced may have the same local properties(T, $\mu_{B}$), yet the volume of the system will vary significantly from event to event\cite{label41}. These fluctuations are to be minimized by introducing suitable fluctuation measures. It has been shown within the model independent sources that such measures may be constructed from the first and second moments of two extensive variables. The first such measure was introduced by Ga$\acute{z}$dzicki and Mrowczynski\cite{label42}. The concept was then generalized\cite{label42} and yet another measure was constructed. The two measures $\Delta$ and $\Sigma$, referred to as the strongly intensive quantities are defined as\cite{label4, label41, label43}
\begin{equation} \label{eq1}
\Delta[A,B] = \frac{1}{C_{\Delta}}[\langle B \rangle \omega[A] - \langle A \rangle \omega[B]]
\end{equation}
and
\begin{equation} \label{eq2}
\Sigma[A,B] = \frac{1}{C_{\Sigma}}[\langle B \rangle \omega[A] + \langle A \rangle \omega[B] - 2 (\langle AB\rangle - \langle A\rangle \langle B\rangle)]
\end{equation}
where, $A$ and $B$ are two extensive variables which are independent of source number distribution. $\omega[A] = \frac{\langle A^2\rangle - \langle A\rangle ^2}{\langle A\rangle}$, $\omega[B] = \frac{\langle B^2\rangle - \langle B\rangle ^2}{\langle B\rangle}$ are scaled variances of A and B, while the quantities within $\langle ...\rangle$ are referred to as the mean values. The normalization factor $C_{\Delta}$ and $C_{\Sigma}$ are required to be proportional to the first moment of any extensive quantity\cite{label45}. The measures $C_{\Delta}$ and $C_{\Sigma}$ are dimensionless and have a common scale, required for a quantitative comparison of fluctuations of different dimensional extensive quantities\cite{label46}. The basic properties of $\Delta[A,B]$ and $\Sigma[A,B]$ are as follows:\\[-0.3cm]
\begin{enumerate}
	\item[1.] In the absence of ebe fluctuations of A and B, $\Delta[A,B]$ = $\Sigma[A,B]$ = 0, i.e., for $\omega[A]$ = $\omega[B]$ = $\langle AB \rangle - \langle A \rangle\langle B \rangle$ =0.
	\item[2.] For independent particle emission, i.e., there is no interparticle correlations, $\Delta[A,B]$ = $\Sigma[A,B]$ = 1. Thus, one can always define $\Delta[A,B] \geqslant$ 0, by exchanging the quantities A and B.
	\item[3.] For model of independent sources {\footnotesize(MIS)}, i.e., when particles are emitted by a number ($n_s$) of identical sources, which are independent of each other and $P(n_s)$ is the distribution of these number, then $\Delta[A,B](n_s)$ and $\Sigma[A,B](n_s)$ are independent of $n_s$ (intensive measures) and of its distribution $P(n_s)$ (strongly intensive measures). Thus, for a model, where {\footnotesize AA} collisions are incoherent superposition of many independent nucleon-nucleon (nn) collisions, $\Delta[A,B]$ and $\Sigma[A,B]$ will be independent of number of sources and hence insensitive to the centrality of the collisions. The two variables will, therefore, assume the same values for {\footnotesize AA} and nn collisions\cite{label46}.
\end{enumerate}

\noindent In the independent particle model the following assumptions are made\cite{label45}.\\[-0.3cm]
\begin{enumerate}
	\item[i)] The state (microstate of the grand canonical ensemble ({\footnotesize GCE}) or a final state of AA collision) quantities A and B are expressed as:
	\begin{eqnarray}
	A = \alpha_1 + \alpha_2 + .. + \alpha_N \quad , \quad B = \beta_1 + \beta_2 + .. + \beta_N  
	\end{eqnarray}
	where, $\alpha_j$ and $\beta_j$ represent the single particle contributions to A and B respectively and N denotes the number of particles.
	\item[ii)] Interparticle correlations are absent. This means that probability of any multiparticle state is the product of probability distributions $\rho(\alpha_j - \beta_j)$ of single particle states and these probability are the same for all $j = 1, 2, .., N$ and independent of N. 
\end{enumerate}
	
\noindent Within IPM the first and the second moments of A and B are given as:
	\begin{eqnarray}
	\langle A\rangle = \bar{\alpha}\langle A\rangle,\quad \langle A^2\rangle = \bar{\alpha^2}\langle N\rangle + \bar{\alpha}^2\left[\langle N^2\rangle - \langle N\rangle\right], \\
	\langle B\rangle = \bar{\beta}\langle B\rangle,\quad \langle B^2\rangle = \bar{\beta^2}\langle N\rangle + \bar{\beta}^2\left[\langle N^2\rangle - \langle N\rangle\right], \\
	\langle AB\rangle = \bar{\alpha\beta}\langle N\rangle + \bar{\alpha}\bar{\beta}\left[\langle N^2\rangle - \langle N\rangle\right]
	\end{eqnarray}
	
\noindent The mean values $\langle A\rangle$ and $\langle B\rangle$ are proportional to the mean number of particles, $\langle N\rangle$ and hence the average size of the system. These quantities are extensive. The quantities $\bar{\alpha}$, $\bar{\beta}$ and $\bar{\alpha^2}$, $\bar{\beta^2}$ and $\bar{\alpha\beta}$ respectively denote the first and second moments of the single particle distribution $P(\alpha,\beta)$. These are independent of $\langle N\rangle$ within IPM and play the role of intensive quantities. It has been shown in ref.44 that the scaled variance $\omega[A]$, which denotes the state-by-state fluctuations of A and mathematically can be expressed as:\\
	\begin{eqnarray}
	\omega[A] = \frac{\langle A^2\rangle - \langle A\rangle ^2}{\langle A\rangle} = \frac{\bar{\alpha^2} - \bar{\alpha}^2}{\bar{\alpha}} = \bar{\alpha} \frac{\langle N^2\rangle - \langle N\rangle ^2}{\langle N\rangle} = \omega[\alpha] + \bar{\alpha} \omega[N] 
	\end{eqnarray}

\noindent where $\omega[\alpha]$ refers to as the scaled variance of single particle quantity $\alpha$ while $\omega[N]$ denotes the scaled variance of N. A similar expression for $\omega[B]$ may also be obtained.\\
	
\noindent Variances $\omega[\alpha]$ and $\omega[\beta]$ depend on the fluctuations of particle number via $\omega[N]$ and hence are not the strongly intensive variables. Using Eqs.4-6 $\Delta[A,B]$ and $\Sigma[A,B]$ may be expressed as:
	\begin{eqnarray}
	\Delta[A,B] = \frac{\langle N\rangle}{C_{\Delta}}\left[\bar{\beta}\omega[\alpha] - \bar{\alpha}\omega[\beta]\right] \\
	\Sigma[A,B] = \frac{\langle N\rangle}{C_{\Sigma}}\left[\bar{\beta}\omega[\alpha] - \bar{\alpha}\omega[\beta] - 2\left(\bar{\alpha}\bar{\beta} - \bar{\alpha}\bar{\beta}\right)\right]
	\end{eqnarray}
	
\noindent Thus, the requirement that $\Delta[A,B] = \Sigma[A,B] =1$ within IPM leads to:
	\begin{eqnarray}
	C_{\Delta} = \langle N\rangle\left[\bar{\beta}\omega[\alpha] - \bar{\alpha}\omega[\beta]\right] \quad and \\
	C_{\Sigma} = \langle N\rangle\left[\bar{\beta}\omega[\alpha] - \bar{\alpha}\omega[\beta] - \left(\bar{\alpha}\bar{\beta} - \bar{\alpha}\bar{\beta}\right)\right]
	\end{eqnarray}
	
\noindent It may be mentioned here that Eqs.4-6 have same structure as Eqs.2-4 of ref.4, obtained within the model of independent sources. The only difference is that the number of sources, $n_s$, in {\footnotesize MIS} is replaced by number of particles, N, in {\footnotesize IPM}. Many particles can be produced from each source and this number would fluctuate from source to source and from event to event. Moreover, particles emitted from one source may be correlated. Therefore, the {\footnotesize MIS}, in general, does not satisfy the assumptions of {\footnotesize IPM}\cite{label45}. Explicit expressions of $\Delta[A,B]$ and $\Sigma[A,B]$ and their normalization factors $C_{\Sigma}$ and $C_{\Delta}$ calculated within {\footnotesize IPM} may be found in ref.44. Calculations performed for the two popular choices of A and B: $i)$ to the study of $p_{\rm T}$ fluctuations using the variables $p_{\rm T}$ and N, $ii)$ chemical fluctuations using variables k and $\pi$ and also for most general case of partially overlapping A and B within the same mathematical formalism. The scaled variance of {\footnotesize MD}, $\omega[N]$ = 0, when N is constant from event to event and $\omega[N]$ = 1 for a Poisson {\footnotesize MD}. The measures $\Delta[p_{\rm T},N]$ and $\Sigma[p_{\rm T},N]$ have similar advantages to $\omega[N]$ because it can be inferred whether the fluctuations are large($>$1) or small($<$1) as compared to independent particle production.

The two measures, $\Delta[p_{\rm T},N]$ and $\Sigma[p_{\rm T},N]$ have been studied\cite{label41, label45} for ideal Bose and Fermi gases within {\footnotesize GCE}. The following general relations have been reported:
\begin{eqnarray}
\Delta^{Bose}[p_{\rm T}, N] < \Delta^{Boltzmann}[p_{\rm T}, N] = 1 < \Delta^{Fermi}[p_{\rm T}, N] \\
\Sigma^{Fermi}[p_{\rm T}, N] < \Sigma^{Boltzmann}[p_{\rm T}, N] = 1 < \Sigma^{Bose}[p_{\rm T}, N]
\end{eqnarray}

\noindent This implies that Bose statistics makes $\Delta[p_{\rm T},N] <$ 1 while $\Sigma[p_{\rm T},N] >$ 1 whereas Fermi statistics works in an opposite manner. Bose statistics gives around 20$\%$ decrease in $\Delta[p_{\rm T},N]$ values and 10$\%$ increase in $\Sigma[p_{\rm T},N]$ values at T = 150 MeV in comparison to {\footnotesize IPM} results, i.e., $\Delta[p_{\rm T},N]$ = $\Sigma[p_{\rm T},N]$ = 1. Modifications in the values of the two measures have been reported\cite{label45} to be rather in-significant for the typical T and $\mu_{B}$. In the model based studies, e.g., {\footnotesize URQMD}, the value of $\Delta[p_{\rm T},N] <$ 1 and $\Sigma[p_{\rm T},N] >$ 1 have been argued to be due to the several sources of fluctuations considered as the model does not include Bose and Fermi statistics. The study of the dependence of $\Delta[p_{\rm T},N]$ and $\Sigma[p_{\rm T},N]$ on the $\eta$ window widths and $p_{\rm T}$ restrictions, for pp collisions at {\footnotesize SPS} and {\footnotesize LHC} energies in the framework of multipomeron exchange model, it has been reported\cite{label47} that $\Delta[p_{\rm T},N]$ and $\Sigma[p_{\rm T},N]$ acquire a values $<$ 1 and $>$ 1 respectively. It has also been suggested that the model can be generalized for pA and AA collisions too\cite{label48} but in that case the enhanced influence of volume fluctuations on $\Delta$ and $\Sigma$ measures can not be ignored rather, must be taken into consideration. It has been shown that as in the case of string fusion\cite{label49}, collective effects considered in this model suggest that these variables would depend on volume fluctuations.\\
	
\noindent The variables $\Delta[A,B]$ and $\Sigma[A,B]$, in most of the fluctuation studies, have been opted as A = $p_{\rm T}$ and B = N. These calculations are based on the consideration of a phase space window of desired range. {\footnotesize NA49} collaboration\cite{label46, label50} has studied ebe fluctuations using $\Delta[p_{\rm T},N]$ and $\Sigma[p_{\rm T},N]$ variables in PbPb collisions at E$_{Lab}$ = 20A, 30A, 40A, 80A and 158A GeV and also for pp, CC, SiSi collisions at E$_{Lab}$ = 158A GeV. It has been reported\cite{label46} that within the considered kinematic range, $\Delta[p_{\rm T},N]$ and $\Sigma[p_{\rm T},N]$ exhibit no significant energy dependence. However, a remarkable system size dependence is observed with the largest values of $\Delta[p_{\rm T},N]$ and $\Sigma[p_{\rm T},N]$, measured in peripheral PbPb collisions. Wu et al\cite{label55} have studied ebe fluctuations using these variables in the framework of {\footnotesize AMPT} and {\footnotesize MIS} models in the energy range, $\sqrt{s_{\rm {NN}}}$ = 7-200 GeV. However, a detailed study of the dependence of these variables on collision centrality and centrality bin width is needed so as to test that to what extent the volume fluctuations may be kept under control by selection of centrality bins of proper widths. By analogy with the $p_{\rm T}-N$ case, several other sets of strongly intensive variables may also be considered. The study of joint fluctuations of two quantities in two $\eta$ windows separated from each other may be of interest too, because they are closely connected with the studies of forward-backward {\footnotesize (FB)} correlations\cite{label5}. {\footnotesize FB} correlations are generally quantified by the correlation coefficient which, incidentally, is not strongly intensive and hence sensitive to the centrality selection. However, using the strongly intensive quantities, trivial fluctuations can be suppressed, which allows to study the intrinsic properties of the particle emitting sources. An attempt has, therefore, been made to study the joint fluctuations of multiplicities and sum of their transverse momenta in two separated $\eta$ windows.

\section{Results and Discussion}

\noindent The present analysis has been carried out by taking into consideration the produced charged particles having their $p_{T}$ values lying in the range $0.005 < p_{T} < 1.5$ GeV and $\eta$ values in the range $\eta = \pm (\eta_{c} + 2.0)$, $\eta_c$ being the center of symmetry of the $\eta$ distribution. The analysis is performed by considering one $\eta$ window and then a pair of $\eta$ windows. In one $\eta$ window analysis, the strongly intensive variables, $\Delta[p_{\rm T}, N]$ and $\Sigma[p_{\rm T}, N]$ are examined as a function of varying width of $\eta$ window within the considered $\eta$ range. This corresponds to changing the rapidity averaged baryon-chemical potential at the freeze-out stage\cite{label50a}. Using two window analysis method, search for short- and long-range correlations is made by examining the dependence of strongly intensive variables $\Sigma[A,B]$ on the separation of the two $\eta$ windows in which variables $A$ and $B$ are evaluated.\\

\subsection{P$_{T}$N fluctuations in $\eta$ window of varying width}
Variable $A$ occurring in Eqs. 1 and 2 is taken as the event multiplicity, whereas $B$ is the scalar sum of $p_{\rm T}$ of all charged particles produced in an event and lying within the considered $\eta$ and $p_{\rm T}$ ranges. The following variables are, thus calculated as:\\

i) Intensive quantity, 
\begin{equation} \label{eq3}
 \omega[N] = \frac{\langle N^2\rangle - \langle N\rangle ^2}{\langle N\rangle} \quad ; \quad
 \omega[p_{\rm T}] = \frac{\langle p_{T}^2\rangle - \langle p_{T}\rangle ^2}{\langle p_{T}\rangle}
 \end{equation} 
where average is taken over all events. \\

ii) Strongly intensive quantities,
\begin{equation} \label{eq4}
\Sigma[p_{\rm T}, N] = \frac{1}{C_{\Sigma}}\left(\langle N\rangle \omega[p_{\rm T}] + \langle p_{\rm T}\rangle \omega[N] - 2(\langle p_{\rm T}N\rangle - \langle p_{\rm T}\rangle \langle N\rangle)\right)
\end{equation}
and
\begin{equation} \label{eq5}
\Delta[p_{\rm T}, N] = \frac{1}{C_{\Delta}}\left(\langle N\rangle \omega[p_{\rm T}] - \langle p_{\rm T}\rangle \omega[N]\right)
\end{equation}
on normalization one may get\cite{label11, label41, label45}
\begin{eqnarray} \label{eq6}
C_{\Delta} = C_{\Sigma} = \langle N\rangle \omega[p_{T}]  \nonumber \\
\omega[p_{\rm T}] = \frac{\bar{p_{\rm T}^2} - \bar{p_{\rm T}}^2}{\bar{p_{\rm T}}} 		\nonumber
\end{eqnarray}
where $\bar{...}$ denotes the inclusive averaged value of particles in all events.\\

\noindent As mentioned earlier, the strongly intensive fluctuation measures, $\Delta$ and $\Sigma$ depend neither on system volume nor on volume fluctuations within the event ensemble. In AA collisions, the volume of the system changes from event to event and such changes can not be eliminated completely. The strongly intensive quantities would permit to overcome such fluctuations, at least partially. Furthermore, the fluctuations in particle multiplicities have significant contributions from statistical or random components. These fluctuations occur due to finite particle multiplicity, centrality selection, limited acceptance of the detector, effect of re-scattering, etc. Contribution to the fluctuations due to changing impact parameter from event to event can be minimized by the proper selection of centrality bin width \cite{label10}. Selection of a very narrow centrality bin would reduce the geometrical fluctuations to its minimum. However, such a narrow centrality bin selection may or may not be possible due to the centrality resolution of the detector. Moreover, a very narrow centrality window would introduce fluctuations due to limited statistics. Hence one has to restrict to somewhat wider centrality windows, e.g., $5\%$ or $10\%$ of the total cross section and then apply the corrections to the inherent fluctuations by taking the weighted mean of the observable, X as,
\begin{eqnarray} 
X = \frac{\Sigma_i n_i X_i}{\Sigma_i n_i}
\end{eqnarray}
where $X_i$ denotes the variable in the $i^{th}$ bin whereas $\Sigma_i n_i = N$ is the total number of events in the $i^{th}$ bin. Values of charged particle multiplicity, N and sum of particle's transverse momenta, $p_{\rm T}$ in a pseudorapidity window of fixed width, \textbf{$\Delta\eta = 1$} are evaluated for various centrality intervals. Here, only those particles are considered which have their $p_{\rm T}$ values lying in the range $\sim$(0.005 -- 1.5) GeV/$c$. The values of collisions centrality and $p_{\rm T}$ range are considered such that the finding may be compared with NA49 results on Pb-Pb collisions in the energy range, $E_{Lab} \simeq$ (20 -- 158A) GeV\cite{label46}. The pseudorapidity window is so chosen that its center coincides with the center of symmetry of the $\eta$ distribution, $\eta_c$. Thus, all the charged particles having their $p_{\rm T}$ values in the considered $p_{\rm T}$ range and $\eta$ values lying in the interval, \textbf{$\eta_c-0.5 < \eta < \eta_c+0.5$} are considered. Collision centrality classes are estimated from the impact parameter dependence on the centrality bins, as described by Klochkov and Selyuzhenkov\cite{label51}.\\

\noindent Shown in Fig.1 (left panel) are the variations of $\Delta[p_{\rm T},N]$ with centrality for 2$\%$, 5$\%$ and 10$\%$ centrality windows. The statistical errors associated with \(\Delta[p_{\rm T}, N]\) are too small to be visible in the figure. The errors are estimated  using the sub-sample method\cite{label52, label53}. The event ensemble is divided into 30 sub-samples and the value of  \(\Delta[p_{\rm T}, N]\) for each sub-sample is calculated individually. Using these values the mean and dispersion are evaluated employing the relations:
\begin{eqnarray} 
\langle \Delta[p_{\rm T},N]\rangle = \frac{1}{n}\Sigma_i \Delta[p_{\rm T},N] \nonumber \\
\sigma = \sqrt{\Sigma_i\frac{(\Delta[p_{\rm T},N]_i - \langle \Delta[p_{\rm T},N]\rangle)^2}{n-1}} \nonumber
\end{eqnarray}

\noindent The statistical error is then calculated as
\begin{eqnarray}
   (Error)_{stat} = \sigma/\sqrt{n} \nonumber
\end{eqnarray}
\noindent where \(n\) denotes the number of sub-samples. It is observed in the figure that for 2$\%$ and 5$\%$ centrality bin the $\Delta[p_{\rm T},N]$ values are independent of centrality bin width. However on considering the centrality bin width as 10$\%$, values of $\Delta[p_{\rm T},N]$ are significantly higher than those obtained for 2$\%$ and 5$\%$ bin widths, particularly for central and semi-central collisions. This difference of $\Delta[p_{\rm T},N]$ values for 10$\%$ and 5$\%$ centrality bin width is rather more pronounced at higher beam energies. It is, however, interesting to note that after applying the bin width correction, $\Delta[p_{\rm T},N]$ values attain constancy irrespective of the widths of the centrality bins, as is evident from the right panel of Fig.1. These observations tend to suggest that while studying $p_{\rm T}$ fluctuations in terms of $\Delta$ variables, the centrality bin should not be as wide as 10$\%$. It is rather safe to deal with 5$\%$ centrality cuts so that the fluctuations arising due to volume fluctuations are under control. It may also be noted from the figure that the values of $\Delta[p_{\rm T},N]$ are the least for most central collisions. These values increase with centrality upto $\sim$ 40$\%$ central collisions and then decrease. It may also be noted that $\Delta[p_{\rm T},N]$ values are $\leqslant$ 1 for 0-5$\%$ central collisions and $>$ 1 for centralities beyond 5$\%$. Thus, $\Delta[p_{\rm T},N]$ values $<$ 1 for most central collisions are smaller than expected from independent particle production. This suggests that in {\footnotesize URQMD}, there are other sources of correlations present which give $\Delta[p_{\rm T},N]$ $<$ 1 as the model does not take into account Bose-Einstein correlation\cite{label46}. These observations are in accord with those reported by NA49 collaboration\cite{label46} for both data as well as models. Moreover, the values of the variable $\Delta[p_{\rm T},N]$ are noted to be slightly less than unity and independent of the beam energy.\\

\noindent Centrality dependence of $\Sigma[p_{\rm T}, N]$ values at the four different beam energies are displayed in Fig.2. It is observed that the data points for 2$\%$, 5$\%$ and 10$\%$ centrality bins overlap and  form a single curve. Calculations carried out in the framework of dipole-based parton string fusion model for PbPb and XeXe collisions at {\footnotesize LHC} energies also suggest that $\Sigma[p_{\rm T}, N]$  values are insensitive to the centrality bin widths. Moreover, nearly the same values of $\Sigma[p_{\rm T}, N]$  against centrality has also been observed in the study by Kovalenko\cite{label61}. {\footnotesize URQMD} simulations have also been carried out by Gorenstein and Grebieszkow\cite{label43} for PbPb collisions at E$_{Lab}$ = 20A GeV to study the effects of centrality selection and limited detector acceptance and efficiency in {\footnotesize AA} collisions in terms of variables $\Delta[p_{\rm T},N^-]$ and $\Sigma[p_{\rm T}, N^-]$; $N^-$ being the multiplicity of the negative hadrons. It has also been reported\cite{label43} that $\Delta[p_{\rm T},N^-]$ is quite sensitive to the centrality bin width, while $\Sigma[p_{\rm T}, N]$ is observed to be nearly insensitive and acquire a value $\sim$ 1 for a centrality bin as narrow as 0-5$\%$ and as wide as 0-20$\%$. In our study, we too observe that $\Sigma[p_{\rm T}, N]$ values are nearly the same irrespective of the centrality bin widths, whereas $\Delta[p_{\rm T},N]$ are observed to be higher for wider centrality bins and acquire similar values after bin width corrections. It may be noted from the figure that $\Sigma[p_{\rm T}, N]$ $<$ 1 for most central collisions for $10A$ GeV data. It increases gradually with collision centrality and ultimately acquires values $>$ 1 for 40$\%$ centrality and beyond. This energy dependence is noted to be rather weaker for 20A GeV data and appear to vanish for energies 30A and 40A GeV, where values of $\Sigma[p_{\rm T}, N]$ are slightly $>$ 1 and nearly constant. These observations tend to suggest that there might be some additional sources of correlations in the model, that start contributing significantly at higher energies.\\

\noindent Values of fluctuation measures, $\Delta[p_{\rm T}, N]$ and $\Sigma[p_{\rm T}, N]$ for the most central(0-7$\%$) collisions are calculated for Au-Au collisions at $E_{Lab} = 10A, 20A, 30A$ and $40A$ GeV. Dependence of $\Delta[p_{\rm T}, N]$ and $\Sigma[p_{\rm T}, N]$ on $\eta$ window widths has been looked into because such a study is expected to provide useful information about the medium created in AA collisions. It would help understand as to how the fluctuations evolve through a purely hadronic medium and hence carves out the region of optimum window size $\Delta\eta_0$ which is best suited for fluctuation studies\cite{label18}. The optimization of $\eta$ window size is important because in small windows, the fluctuations may be affected by global charge conservation and some other aspects may not be visible, whereas if $\eta$ window is too large, it may contain more number of particles and fluctuations may diffuse\cite{label18}. Moreover, since the URQMD model takes care of several sources of fluctuations and correlations and resonance decay\cite{label14, label18, label21}, therefore, while searching for resonance decays, $\eta$ window width is to be chosen carefully to capture the correlated hadrons resulting from the decay of the resonances\cite{label18, label54}. The two hadrons resulted from the decay of resonances are expected to have a pseudorapidity difference of the order of unity. Therefore, to enlarge the probability of simultaneous hits of both the correlated hadrons, e.g., \(\pi^+\pi^-\) from the resonance decays, a window of width \(\Delta\eta \ge 1\) should be chosen. Thus, the \(\eta\) window width should be large enough, however, this \(\Delta\eta\) should be small in comparison to the entire \(\eta\) interval \(\Delta\eta (\simeq ln\sqrt{S_{NN}}/m)\), where m is the mass of the hadron\cite{label54}.  The values of $\Delta[p_{\rm T}, N]$ and $\Sigma[p_{\rm T}, N]$ are calculated by selecting a $\eta$ window of width 0.4 units and placed in such a way that its center coincides with the center of symmetry of $\eta$ distribution, $\eta_c$. The window width is then increased in steps of 0.4 units until the region $\Delta\eta$ = 4.0 is covered. Variations of $\Delta[p_{\rm T}, N]$ and $\Sigma[p_{\rm T}, N]$ with $\Delta\eta$ are shown in Fig.3. It may be seen in the figure that $\Delta[p_{\rm T}, N] \sim$ 1 for $\Delta\eta$ = 0.4. It then increases with increasing $\Delta\eta$(upto $\Delta\eta \simeq$ 1.6) and then decreases to zero. $\Sigma[p_{\rm T}, N]$ values, for 10A GeV data, are noticed to decrease first with increasing $\Delta\eta$ and then acquire a saturation. However, for 20A, 30A and 40A GeV data the values of $\Sigma[p_{\rm T}, N]$ slightly increase and then it decrease. It should be mentioned here that the variations in the $\Sigma[p_{\rm T}, N]$ values are within 5$\%$ from unity for all the data sets. The difference in the observed trend of variations of $\Sigma[p_{\rm T}, N]$ for 10A GeV data as compared to the other data sets might be due to the fact that in the model at lower energies($E_{Lab} <$ 12.4A GeV) the system changes from baryon dominated system to meson dominated system as one moves from E$_{Lab}$ = 12A GeV to higher energies. At energies below $E_{Lab}$ = 12.4A GeV($\sqrt{s} <$ 5 GeV) the model accounts for known hadrons and resonances while for beam energies $\sqrt{s} >$ 5 GeV the excitation of color string and their subsequent fragmentation into hadrons are considered\cite{label14, label18}. It may also be noted in the figure that $\Sigma[p_{\rm T}, N]$ values are less affected by the widths of $\eta$ windows as compared to $\Delta[p_{\rm T}, N]$ values. This indicates that the measure $\Delta[p_{\rm T}, N]$ is rather more sensitive to the limited detector acceptance and efficiency as compared to $\Sigma[p_{\rm T}, N]$. Moreover, since the two measures depend on the correlation length too, therefore, when the kinematic acceptance is much smaller than the correlation range, the effect will be washed out\cite{label43}. Hence, while comparing the experimental findings with the model predictions kinematical restrictions should be addressed carefully. \\

\noindent The strongly intensive quantities have a characteristic property\cite{label45, label55}, that their values are equal to one for independent emission of particles. This is the assumption of independent particle model(IPM)\cite{label54}. Events corresponding to IPM may be simulated by event mixing technique\cite{label45,label55,label56}. There are a variety of mixed event models\cite{label55}. The one adopted in the present study, gives results identical to the IPM in the limit of infinite number of original and mixed events. A mixed event of the same multiplicity is simulated by randomly selecting its N particles from different events and combining them to get this artificial events. Thus, the multiplicity distributions of original and mixed events samples would be identical but will be deprived of the inter-particle correlations. Consequently, the mixed event ensemble would satisfy the IPM ensemble\cite{label56}, i.e., for such samples, $\Delta[p_{\rm T}, N]$ = $\Sigma[p_{\rm T}, N]$ = 1. Variations of $\Delta[p_{\rm T}, N]$ and $\Sigma[p_{\rm T}, N]$ with $\Delta\eta$ for mixed event sample are also displayed in Fig.3. It is interesting to note in the figure that $\Delta[p_{\rm T}, N]$ and $\Sigma[p_{\rm T}, N]$ acquire a value $\simeq$ 1, irrespective of $\eta$ window widths, as is expected from IPM model predictions. It may also be observed from the figure(top panels) that for any given $\Delta\eta$, $\Delta[p_{\rm T}, N]$ and $\Sigma[p_{\rm T}, N]$ increase with beam energy. The magnitudes of enhancement in the values of the two quantities with increasing energies appear to become smaller. This might be attributed to the fact that for a given $\Delta\eta$ same $p_{\rm T}$ cuts are applied to all energies but the $\eta$ and $p_{\rm T}$ distributions are quite different at different energies. It has been reported\cite{label55} that $\Delta[p_{\rm T}, N]$ and $\Sigma[p_{\rm T}, N]$ values have different $\eta$ and $p_{\rm T}$ dependence at different energies. The findings based on mixed events analyses, however, do not show any energy dependence. This result, thus, tend to indicate that the trends of variations of $\Delta[p_{\rm T}, N]$ and $\Sigma[p_{\rm T}, N]$ with $\Delta\eta$ are due to the various intricate origins, i.e., resonance decay, $p_{\rm T}$ vs N anti-correlations, taken care of by {\footnotesize URQMD} model.\\

\subsection{$NN$, $Np_{\rm T}$ and $p_{\rm T}p_{\rm T}$ fluctuations in two pseudorapidity windows}
\noindent In analogy with the $[p_{\rm T},N]$ fluctuation studied in the previous section, joint fluctuations of two variables in two distinct $\eta$ windows separated from each other may be studied\cite{label5}. Such studies are regarded as of special interest as they are connected to the investigations of forward-backward({\footnotesize FB}) correlations\cite{label57, label58, label59, label60}. {\footnotesize FB} correlations are usually investigated\cite{label59,label60} in terms of correlation coefficient, which incidentally is not an intensive quantity and depends on the centrality selection in AA collisions\cite{label5,label59,label60}. {\footnotesize FB} correlations, if studied in terms of strongly intensive quantities, might permit to examine the intrinsic properties of particle emitting sources because the trivial fluctuations would be suppressed\cite{label5}. \\

\noindent Strongly intensive measures used in the present study to examine the joint fluctuations of two variables are defined as\cite{label5, label11}
\begin{equation} \label{eq8}
\fl \Sigma[N_{\rm F},N_{\rm B}] = \left(\frac{1}{\langle N_{\rm F}\rangle + \langle N_{\rm B}\rangle}\right) \left(\langle N_{\rm B}\rangle\omega_{N_{\rm F}} + \langle N_{\rm F}\rangle\omega_{N_{\rm B}} - 2(\langle N_{\rm F}N_{\rm B}\rangle - \langle N_{\rm F}\rangle\langle N_{\rm B}\rangle \right)
\end{equation}
\begin{eqnarray} \label{eq9}
\fl \Sigma[N_{\rm F},p_{\rm T_B}] = \left(\frac{1}{(\langle N_{\rm F}\rangle + \langle N_{\rm B}\rangle)\bar{p_{\rm T}} + \langle N_{\rm F}\rangle \omega_{p_{\rm T_B}}}\right) & \nonumber \\ \left( \langle p_{\rm T_B}\rangle \omega_{N_{\rm F}} + \langle N_{\rm F}\rangle \omega_{p_{\rm T_B}} - 2(\langle N_{\rm F}p_{\rm T_B}\rangle - \langle N_{\rm F}\rangle \langle p_{\rm T_B}\rangle \right)
\end{eqnarray}

\begin{eqnarray} \label{eq10}
\fl \Sigma[p_{\rm T_F},p_{\rm T_B}] = \left(\frac{1}{\langle p_{\rm T_B}\rangle \left(\bar{p_{\rm T_F}} + \omega_{p_{\rm T_F}} \right) + \langle p_{\rm T_F}\rangle \left( \bar{p_{\rm T_B}} + \omega_{p_{\rm T_B}} \right)} \right) & \nonumber \\
\left( \langle p_{\rm T_B}\rangle \omega_{p_{\rm T_F}} + \langle p_{\rm T_F}\rangle \omega_{p_{\rm T_B}} - 2 \left( \langle p_{\rm T_F}p_{\rm T_B}\rangle - \langle p_{\rm T_F}\rangle \langle p_{\rm T_B}\rangle \right) \right)
\end{eqnarray}
where, $N_{\rm F}$, $N_{\rm B}$, $p_{\rm T_F}$ and $p_{\rm T_B}$ respectively denote event-wise particle multiplicities and sum of their transverse momenta in the {\footnotesize F} and $B$, $\eta$ regions respectively, while the quantities with overline $\bar{...}$ represent the event averaged values over the entire data sample. $\bar{p_{\rm T}} = 1/N_{track} \sum_{tracks} p_{\rm T}$ is the mean transverse momentum of all tracks in all events in {\footnotesize F} or $B$ windows. $\omega_{p_{\rm T}}(F,B)$ is the scaled variance of inclusive $p_{\rm T}$ spectra estimated for {\footnotesize F} or $B$ windows. It should be mentioned here that instead of considering event-wise average $p_{\rm T}$ values a sum of event $p_{\rm T}$ values is used to construct strongly intensive quantities for the two variables A and B which should be extensive quantities\cite{label4, label5}. \\

\noindent $\Sigma[A,B]$ for the three pairs of variables defined by Eqs. 8-10 in two $\eta$ windows are calculated for 2$\%$, 5$\%$ and 10$\%$ centrality bins at four incident energies. Two $\eta$-windows, each of width $\Delta\eta = $ 0.5 are considered and placed adjacent to each other such that one lies in the forward and another in the backward regions of $\eta$ distribution. Positions of these windows are thus symmetric with respect to center of symmetry of the $\eta$ distribution. Fig.4 exhibits the centrality dependence of $\Sigma[N_{\rm F}, N_{\rm B}]$ at the four incident energies. It may be noted from the figure that $\Sigma[N_{\rm F}, N_{\rm B}]$ values are independent of the centrality bin widths. This variable acquires a value $\simeq 1$ in 0 to $\sim 60\%$ centrality range. It then rises to a maximum value at $\sim 80\%$ centrality and later tends to decrease for still higher centralities. It has been suggested that the observed maxima in the regions of very peripheral collisions may be related to the transition into a region of diffusive edges of nuclei. But this centrality range is hardly accessible in experiments\cite{label61}. Presented in Figs.5 and 6 are the variations of $\Sigma[N_{\rm F}, p_{\rm T_B}]$ and $\Sigma[p_{\rm T_F}, p_{\rm T_B}]$ values with collisions centrality for the three selections of centrality bin widths. It is noticed in this figure that $\Sigma$ for both pairs of variables exhibits centrality bin width dependence for central and semi-central collisions. The observed centrality bin width dependence, however, vanishes after applying bin width corrections, as shown in right panels of Figs.5 and 6. It may also be noted that for a given centrality class the values of $\Sigma[N_{\rm F}, p_{\rm T_B}]$ are slightly higher than those for $\Sigma[p_{\rm T_F}, p_{\rm T_B}]$ but the trends of variations of $\Sigma$ for both pairs are nearly the same. $\Sigma$ values are observed to decrease first with increasing centrality upto $\sim$ 70$\%$ of central collisions and thereafter, it gradually increase. However, after bin width corrections, the values of this measure show a gradual decrease in its value upto $\sim$ 70$\%$ centrality. Beyond this centrality the values are noticed to show an increasing trend. Almost similar dependence of $\Sigma[N_{\rm F}, N_{\rm B}]$, $\Sigma[N_{\rm F}, p_{\rm T_B}]$ and $\Sigma[p_{\rm T_F}, p_{\rm T_B}]$ have been observed in {\footnotesize MC} studies in the framework of model with string fusion for Pb-Pb collisions at $\sqrt{s_{NN}}$ = 2.76 TeV\cite{label61}.\\

\noindent The study of forward-backward correlations permits to decouple the short-range correlations({\footnotesize SRC}) from the long-range-correlations({\footnotesize LRC})\cite{label62, label63, label64}. The  {\footnotesize SRC} are expected to arise due to the decays of resonances or clusters, jet or mini-jet induced correlations. Contributions from {\footnotesize SRC} are confined to a pseudorapidity region of \(\sim \) two units, whereas, {\footnotesize LRC}, which are envisaged to extend over a rather longer range (\(>\) two units of \(\eta\)), originate from the fluctuations associated with the particle emitting sources, such as strings, clusters, cut pomerons, mini-jets, etc\cite{label62, label63, label64, label65, label66}. Therefore, in order to look for the forward-backward correlations, not only the selection of pseudorapidity window of proper width is important but their separation from each other in \(\eta\) space is equally important. By selecting a small separation between the two \(\eta\) windows {\footnotesize SRC} can be studied while putting a large separation, contributions from {\footnotesize SRC} can be minimized and features of {\footnotesize LRC}  may be examined. Values of $\Sigma[N_{\rm F},N_{\rm B}]$, $\Sigma[N_{\rm F}, p_{\rm T_B}]$ and $\Sigma[p_{\rm T_F}, p_{\rm T_B}]$ are calculated for 7$\%$ central Au-Au collisions at various energies. Two $\eta$ windows, each of width, $\Delta\eta$ = 0.2 are placed adjacent to each other such that they are symmetric with respect to $\eta_c$. Afterwards, the values of $\Sigma[A,B]$ for the three pairs of variables are estimated by counting the particles and calculating the sum of their $p_{\rm T}$ values in the forward(F) and backward(B) windows. The two windows are then shifted in their respective regions, each in steps of 0.2 units of $\eta$ such that their separations($\eta_{sep}$) become 0.4, 0.8, 1.2, etc., until the region, $\Delta\eta \pm$ 2.0 is exhausted. Variations of $\Sigma[A,B]$ for three pairs of variables with $\eta_{sep}$ are shown in Fig.7. It may be noted from the figures that $\Sigma[N_{\rm F},N_{\rm B}]$ values are $\simeq$ 1 for $\eta_{sep} <$ 1 for all the four beam energies considered. {\footnotesize PYTHIA} model calculations carried out for pp collisions at {\footnotesize LHC} energies, 0.9, 2.76 and 7 TeV also indicate that $\Sigma[N_{\rm F},N_{\rm B}] \sim$ 1 for small $\eta$ window separations\cite{label67}. As this separation increases the values of $\Sigma[N_{\rm F},N_{\rm B}]$ increase upto certain saturation points and then gradually decrease. The point of saturation has also been observed to increase with increasing collision energies\cite{label67}. In context to the string model predictions the reduction in $\Sigma[N_{\rm F},N_{\rm B}]$ values are envisaged to occur as a consequence of decrease in the values of $\langle N_{\rm F}\rangle$ and $\langle N_{\rm B}\rangle$ due to the reduction of number of strings contributing to both observation windows\cite{label67}. Variations of $\Sigma[N_{\rm F},N_{\rm B}]$ with $\eta_{sep}$ for the mixed events are also displayed in Fig.7. It is interesting to note that the mixed events sample gives the value of the variable as $\sim$ 1 irrespective of $\eta_{sep}$ values, which is in accord with the {\footnotesize IPM} predictions.\\

\noindent On replacing multiplicity with transverse momenta in one or both the $\eta$ windows, it is noted that $\Sigma[N_{\rm F},p_{\rm T_B}]$ and $\Sigma[p_{\rm T_F}, p_{\rm T_B}]$ first increase with increasing separations between the F and B $\eta$ windows in almost identical fashion and then tend to acquire saturations for values of $\eta_{sep} \geqslant$ 2.8. It may also be noted that for a given $\eta_{sep}$, $\Sigma[p_{\rm T_F}, p_{\rm T_B}]$ values are somewhat smaller as compared to those of $\Sigma[N_{\rm F},p_{\rm T_B}]$. For 0-8$\%$ central Be-Be collisions at 150A GeV, $\Sigma$ values for the three pairs of variables have been observed to increase with increasing $\eta_{sep}$\cite{label5, label8}. These observations appear to be consistent with the predictions of EPOS-1.99 model\cite{label5} and quark-gluon string model for pp interactions at LHC energies\cite{label62}. Some energy dependence of $\Sigma[N_{\rm F},p_{\rm T_B}]$ and $\Sigma[p_{\rm T_F}, p_{\rm T_B}]$ values for relatively smaller separations between the two $\eta$ windows may also be noticed in Fig.7 which decreases as the $\eta_{sep}$ increases and vanishes for $\eta_{sep} \geqslant 2.8$. For the mixed events, however, $\Sigma[N_{\rm F},p_{\rm T_B}]$ and $\Sigma[p_{\rm T_F}, p_{\rm T_B}]$ values are nearly $\simeq$ 0.7 and 0.6 respectively irrespective of $\eta_{sep}$ and beam energy.\\

\noindent Joint fluctuations of multiplicity in two $\eta$ windows have earlier been studied by constructing an event-wise observable C, defined as\cite{label5, label68}
\begin{eqnarray} \label{eq11}
C = \frac{(N_{\rm F} - N_{\rm B})}{\sqrt{N_{\rm F}N_{\rm B}}}
\end{eqnarray}
The variance of C, $\sigma_{c}^{2}$ for a given set of events has the properties similar to that of $\Sigma[N_{\rm F},N_{\rm B}]$ although it is not a strongly intensive quantity\cite{label5}. PHOBOS Collaboration has observed that for Au-Au collisions at $\sqrt{s_{NN}}$ = 200A GeV, $\sigma_{c}^{2} \simeq 1$ for $\eta$ = 0 and somewhat higher on shifting the $\eta$ windows\cite{label69, label70}. These observations agree with what has been reported for 150A GeV Be-Be collisions\cite{label5}. It has been argued\cite{label69,label70} that if {\footnotesize {QGP}} formation takes place near mid-rapidity, any sort of cluster structure, if present, would break, as has been reported for pp collisions\cite{label69,label70,label71}. This would lead to reduction in $\sigma_{c}^{2}$ values. However, the reduction in $\sigma_{c}^{2}$ may also be observed due to clusters produced at $\eta$ = 0 that emit particles in both forward and backward regions including an `intrinsic' {\footnotesize LRC} that decreases $\sigma_{c}^{2}$. With the normalization of $\Delta[A,B]$ and $\Sigma[A,B]$ as proposed\cite{label45}, these quantities become dimensionless thereby, facilitating a quantitative comparison of fluctuations of different extensive quantities\cite{label16}. \\

\noindent In the study carried out in the frame work of dipole-based proton-string model with string fusion it has been observed that in Pb-Pb collisions at $\sqrt{s_{NN}}$ = 2.76 TeV, \(\Sigma[N_F, p{T_B}]\),  \(\Sigma[N_F, N_B]\) and \(\Sigma[p_{T_F}, p_{T_B}]\) are independent of centrality and centrality bin width, whereas in our case (at lower energies) {\footnotesize URQMD} predicts some centrality bin width dependence in central collisions. However the range of \(\Sigma[N_F, p{T_B}]\) and \(\Sigma[p_{T_F}, p_{T_B}]\) obtained in this study are close to the one obtained by us. The two families of strongly intensive measures $\Delta$ and $\Sigma$ has been addressed by NA49 and NA61/SHINE collaborations\cite{label49, label52, label56} for pp and {\footnotesize AA} collisions and compared with {\footnotesize EPOS} and {\footnotesize HSD} model. It is observed\cite{label51} that the {\footnotesize URQMD} predictions are also quite close to those observed for 7.2$\%$ Pb-Pb collisions at different energies. {\footnotesize HSD}, however, overestimates the experimental results. The measures $\Delta(p_{T}, N)$ and $\Sigma(p_{T}, N)$ have also been studied using pp, CC and Cu-Cu collisions in the energy range $\sqrt{s_{NN}}$ = 7.7 to 200 GeV using the event simulated in the framework of {\footnotesize AMPT} model\cite{label55}. Combined event method and the scaled strongly intensive quantities are introduced to partially eliminate the trivial correlations arising due to nucleon-nucleon pairs it is found that system size dependence observed in central collisions almost vanish for peripheral collisions. Studies based on different phenomenological models and analytical calculations considering the different versions of model independent sources: with fixed number of sources and sources having Poisson or negative binomial distributions indicate that the quantities $\Delta(p_{T}, N)$ and $\Sigma(p_{T}, N)$ are independent of mean number of sources and its fluctuations. This reflects that the strongly intensive properties of $\Delta$ and $\Sigma$ measures and is a main motivation of their using for examining the ebe fluctuations in {\footnotesize AA} collisions\cite{label4}.\\

\noindent Besides the $p_{T}$ and $N$, other variables like [k$\pi$] in one window analysis and $N_FN_B$, $N_Fp_{T_B}$ and $p_{T_F}p_{T_B}$ can also be used to look for the validity of model independent sources role of experimental acceptance, centrality selection, etc. However, the studies carried out so far in different transport models, like, {\footnotesize HSD}, {\footnotesize PHSD}, {\footnotesize URQMD} do not show any sign of transition to deconfine phase in fluctuation measures, although the partonic degree of freedom are present in {\footnotesize PHSD} model and the horn in k$^+$/$\pi^+$ ratio is reproduced\cite{label3}.

\noindent In order to look for the energy and system size dependence for NA61/SHINE experiments at SPS, URQMD simulations have been carried out by Grebieszkow\cite{label50a} Event samples corresponding to Be-Be, Ar-Ca, Xe-La collisions at 13A, 20A, 30A,40A, 80A and 158A GeV beam energies were generated and analyzed. It has been reported that the observed system size dependence of \(\Delta\) and \(\Sigma\) measures, for all charged particles, have contributions from these measures due to positively and negatively charged particles and hence the overall contributions may be masked by some trivial sources of fluctuations and correlations arising due to the presence of protons in the sample. Moreover, the correlations between positively and negatively charged particles ( due to particle decays) will also be present.\\

\section{Conclusion}
Dependence of strongly intensive fluctuation measures of transverse momentum and multiplicity on the collision centrality, centrality bin widths and pseudorapidity are studied in Au-Au collisions at FAIR energies. The events corresponding to $E_{Lab}$ = 10A, 20A, 30A and 40A GeV are simulated using the Monte Carlo model URQMD. The findings are compared with the experimental results as well as with the predictions of various {\footnotesize MC} models reported earlier. It is observed that the strongly intensive measures, $\Delta[p_{\rm T},N]$ and $\Sigma[p_{\rm T},N]$ have advantages over the other fluctuation measures. These measures are not too sensitive to the trivial system size fluctuations. The dependence of the two measures on the collision centrality bin is rather moderate and reflects changes in local physical properties for different centrality samples. The findings reveal that the two measures serve as a better tool to look for the effects of kinematical acceptance. $\Delta[p_{\rm T},N]$ reflects a moderate effect of limited acceptance windows on its values, whereas in the case of $\Sigma[p_{\rm T},N]$ these effects are observed to be nearly absent. Furthermore, $\Delta[p_{\rm T},N]$ appears to be more sensitive to the interparticle correlations as compared to $\Sigma[p_{\rm T},N]$ because of its larger deviations from unity in comparison to the $\Sigma[p_{\rm T},N]$ where the deviations from unity are rather small.\\

\noindent The study of joint fluctuations of two variables in two separate \(\eta\) windows is of special interest as such studies are closely related to the investigations involving F-B correlations which have been addressed since last several decades . F-B correlations are usually studied in terms of correlation coefficient, which is not an intensive quantity and strongly depend on the centrality bin width selections. Using the intensive variable, \(\Sigma\), trivial fluctuations may be suppressed and hence information on the intrinsic properties of particle emitting sources may be extracted. The present study reveals that \(\Sigma[N_F, N_B]\) remains unaffected  on the choice of the centrality bin width. For most central collisions, the variable acquires a value \(\sim 1\), as long as the separation between the two \(\eta\) windows is not loo large, i.e., \(\eta < 2\), which might be due to the fact that correlations, if any,  would be suppressed if the kinematical acceptance is smaller than the correlation range. For larger separations, however, \(\Sigma[N_F, N_B]\) values are found to be greater than unity and grows with increasing beam energies. Such an increase in the \(\Sigma[N_F, N_B]\) values is expected to be due to the correlated hadrons resulting from the resonance decays which are captured by the \(\eta\) windows positioned with such separations. Furthermore, the observed decrease in the \(\Sigma[N_F, N_B]\) values for still larger \(\eta_{sep}\) might be because of the smaller number of particles hitting the \(\eta\) windows having positions in the region of \(\eta\) distributions where particle densities go on decreasing quickly. In contrast, \(\Sigma[N_F, N_B]\), \(\Sigma[N_F, p{T_B}]\) and \(\Sigma[p_{T_F}, p_{T_B}]\) exhibit sensitivity to the centrality bin widths and require bin width corrections. This reflects the changes in the local physical properties for different centrality samples.\\

\noindent Some aspects of strongly intensive measures, studied in the framework of URQMD model at FAIR energies lead to some interesting findings, like centrality dependence, centrality bin width effects, detector acceptance, etc. For example, in one window analysis, $\Sigma[p_{\rm T},N]$ is observed to be rather insensitive to the centrality bin width as compared to $\Delta[p_{\rm T},N]$ and hence may be taken as a better choice while dealing with the data from experiments where centrality restrictions are not too narrow. Furthermore, both the measures exhibit beam energy and detector acceptance dependence, although relatively larger deviations from the IPM predictions are observed in the case of $\Delta[p_{\rm T},N]$. This suggests that either the centrality bin $\sim$ 7\% is not too narrow (for $\Delta[p_{\rm T},N]$) to suppress the fluctuations due to changing impact parameter on ebe basis or there might be other sources of correlations present. Protons, may contribute significantly to such correlations. The  $\Sigma$ measure, for such correlations, has been suggested to be much more sensitive~\cite{label72}. The study of joint fluctuations of two variables, multiplicity and transverse momentum, in separated pseudorapidity windows also tend to suggest that the {\footnotesize URQMD} model predicts, although small but systematic deviations from {\footnotesize IPM} predictions at {\footnotesize FAIR} energies. The findings also reveal that choice of the pair of variables is also important.  A complete understanding of all possible sources of fluctuations and correlations in {\footnotesize URQMD}, however, requires more detailed investigations. For  studying the exotic effects, like CP, onset of deconfinement, etc., one should find good reference values of these measures, for which origin of all possible sources of fluctuations and correlations in {\footnotesize URQMD} should be completely understood. This aspect requires somewhat more detailed investigation. The present study, however, gives an idea of the sensitivity of the two measures to  centrality bin width,  limited detector/kinematical acceptance, choice of pairs of variables (for joint fluctuations of two variables), etc. 

\newpage
\centering

\begin{figure}
\begin{center}
	\includegraphics[scale=0.7]{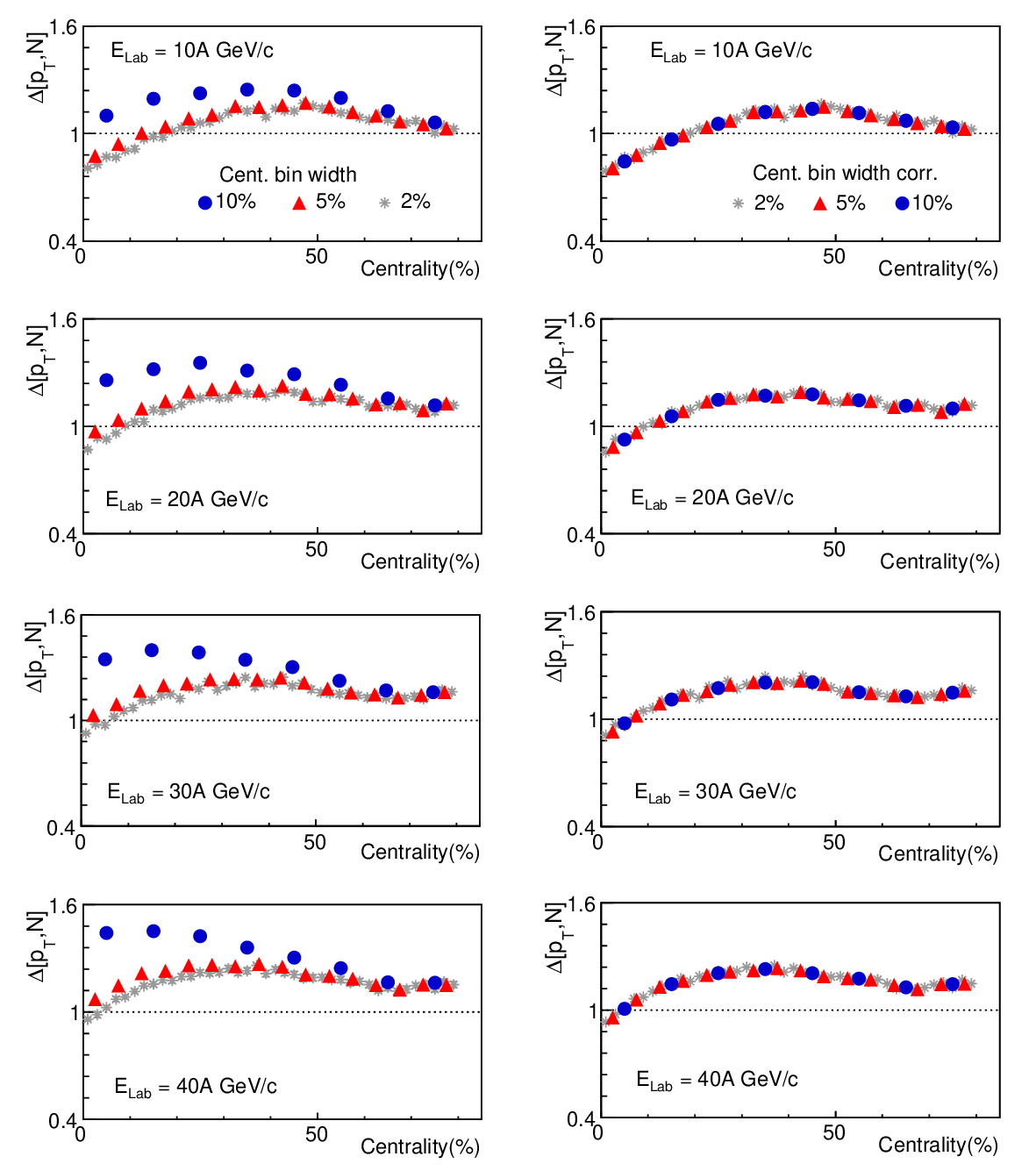}
		\caption{Variations of $\Delta[p_{\rm T},N]$ with collision centrality for $2\%$, $5\%$ and $10\%$ centrality bins at $E_{Lab}$ = 10A, 20A, 30A and 40A GeV.}	
			
\end{center}
\end{figure}

\begin{figure}
\begin{center}

	\includegraphics[scale=0.7]{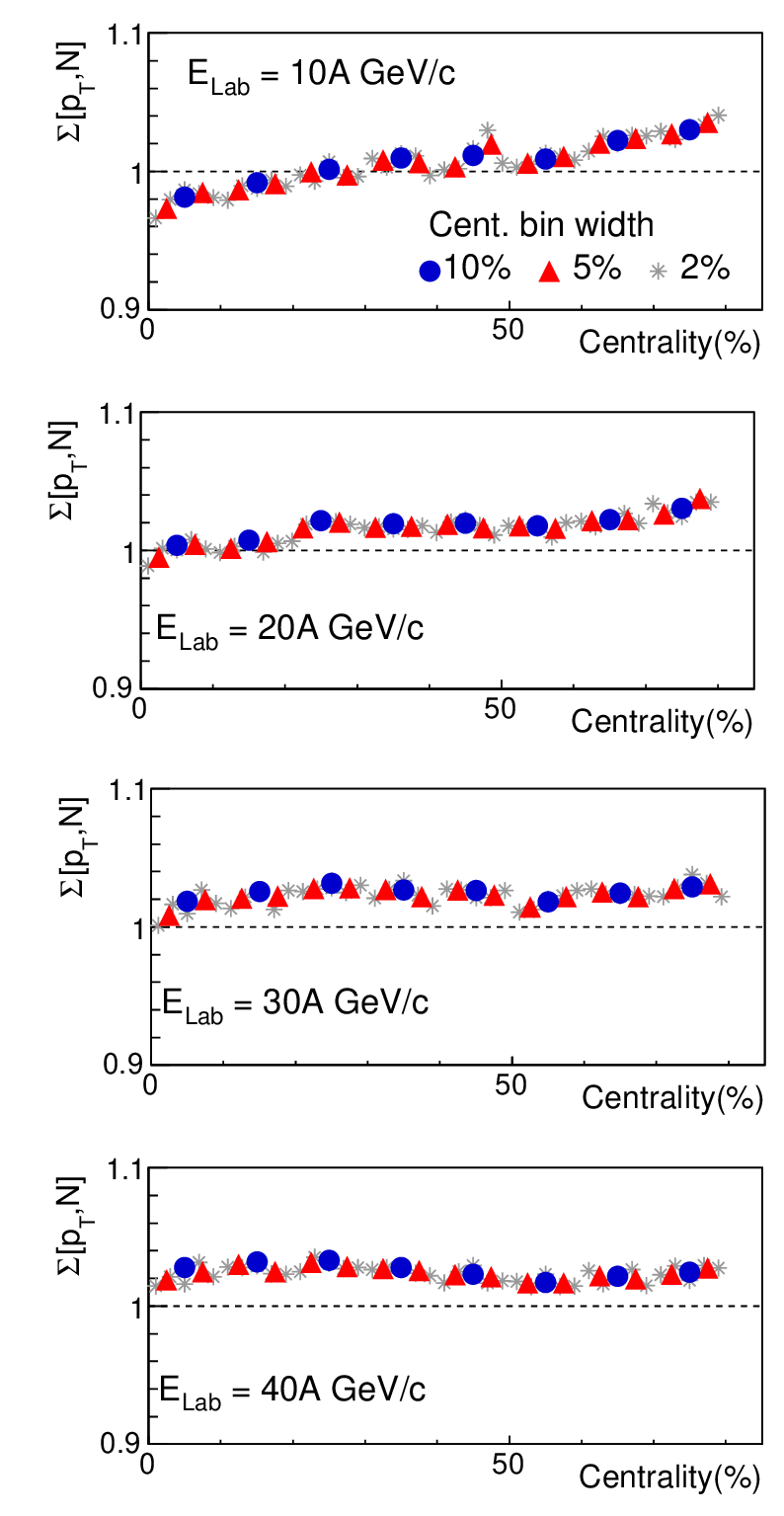}
		\caption{Centrality dependence of $\Sigma[p_{\rm T},N]$ for $2\%$, $5\%$ and $10\%$ centrality bin widths at $E_{Lab}$ = 10A, 20A, 30A and 40A GeV.}	
\end{center}			
\end{figure}

\begin{figure}
\begin{center}
	\includegraphics[scale=0.7]{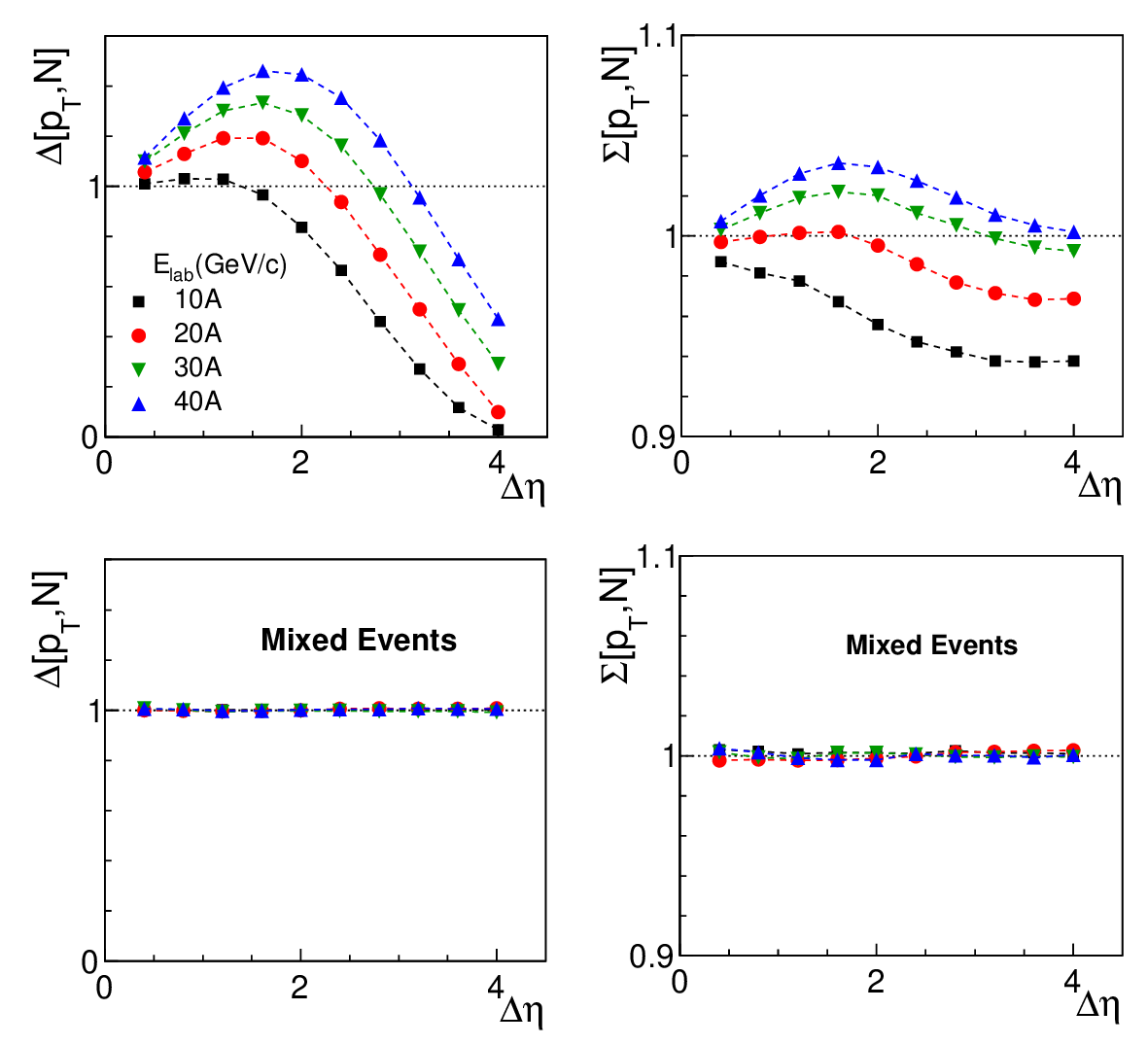}
			\caption{Variations of $\Delta[p_{\rm T},N]$ and $\Sigma[p_{\rm T},N]$ on $\eta$ window width in centrality bin 7$\%$ at $E_{Lab}$ = 10A, 20A, 30A and 40A GeV. Bottom plots show the variations in mixed events.}
\end{center}				
\end{figure}

\begin{figure}
\begin{center}
	\includegraphics[scale=0.7]{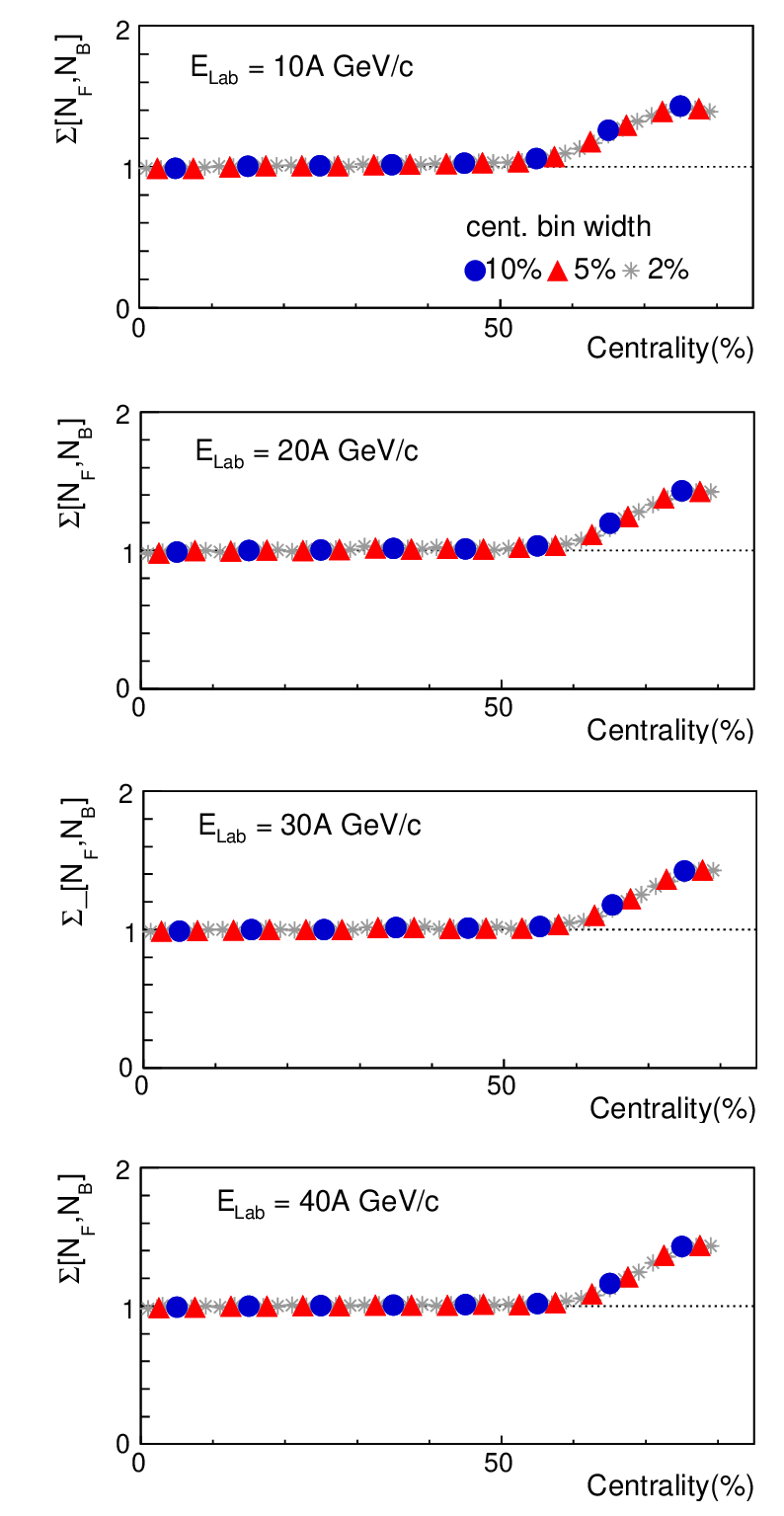}
			\caption{Centrality dependence of $\Sigma[N_{\rm F},N_{\rm B}]$ for centrality bin widths, 2$\%$, 5$\%$ and 10$\%$ at four incident energies.}
\end{center}				
\end{figure}

\begin{figure}
\begin{center}
	\includegraphics[scale=0.7]{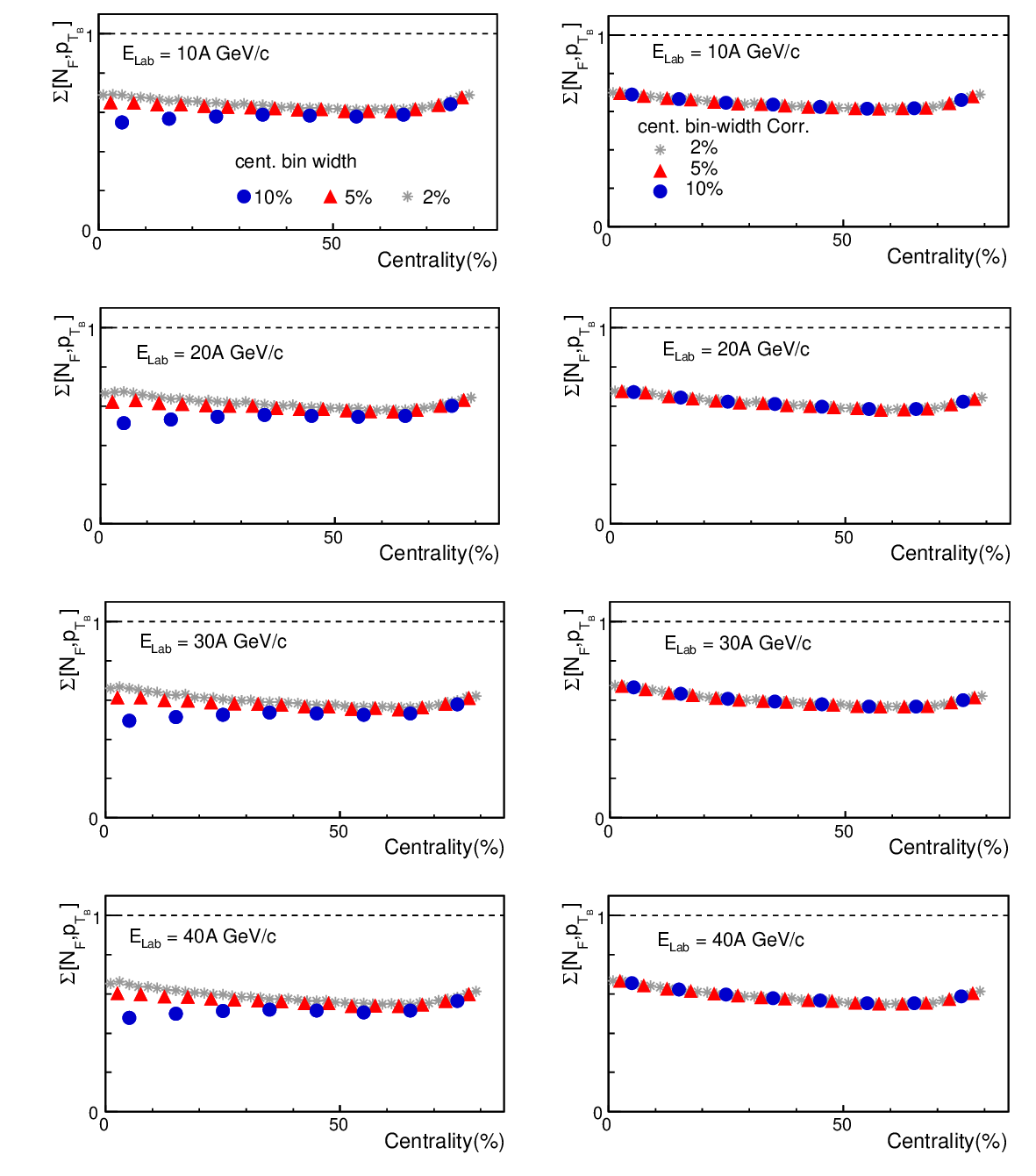}
			\caption{Variations of $\Sigma[N_{\rm F},p_{\rm T_B}]$ with collision centrality for centrality bin widths, 2$\%$, 5$\%$ and 10$\%$.}	
\end{center}							
\end{figure}

\begin{figure}
\begin{center}
	\includegraphics[scale=0.7]{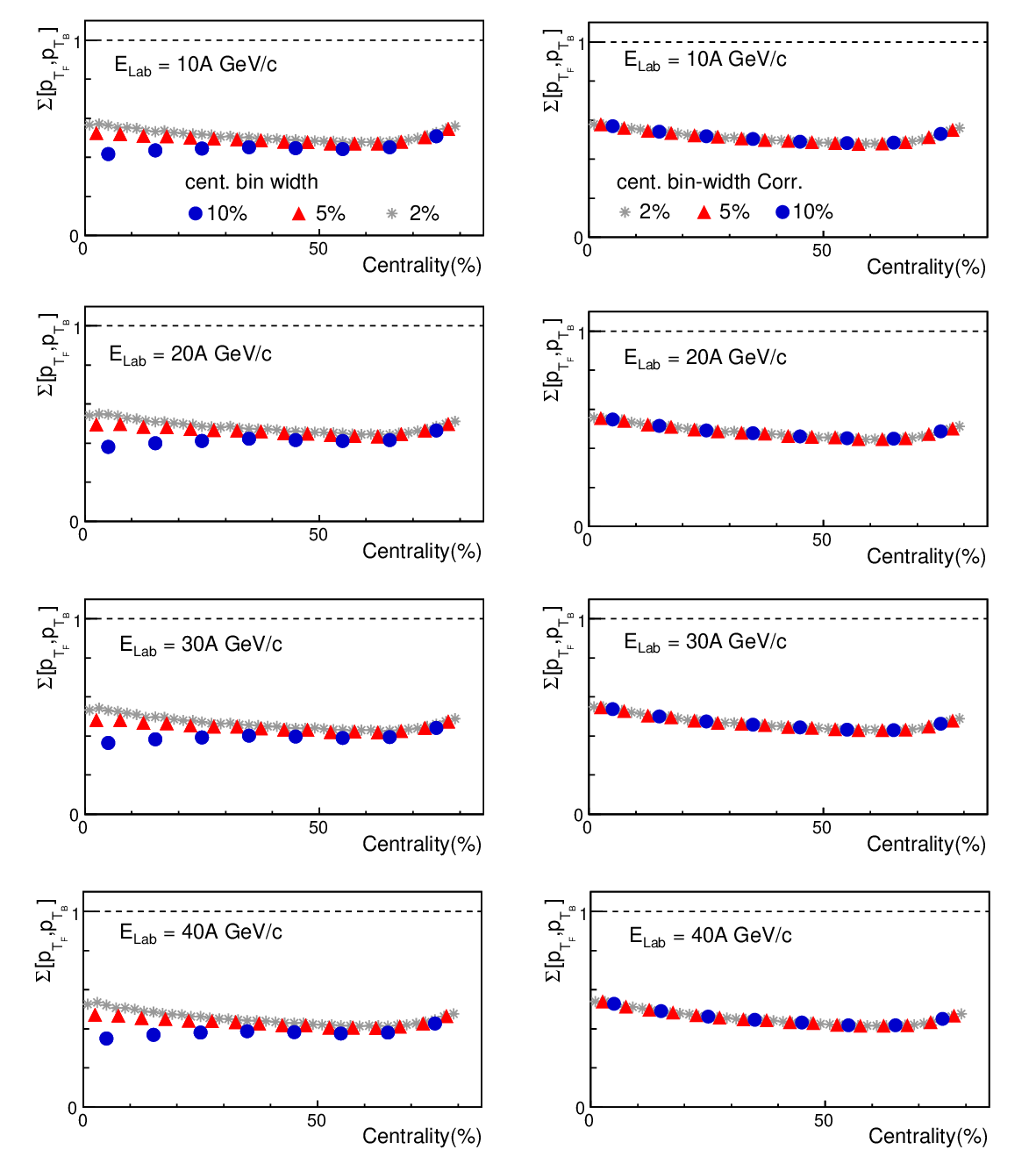}
			\caption{Dependence of $\Sigma[p_{\rm T_F},p_{\rm T_B}]$ on centrality for centrality bin widths, 2$\%$, 5$\%$ and 10$\%$.}	
\end{center}							
\end{figure}

\begin{figure}
\begin{center}
	\includegraphics[scale=0.7]{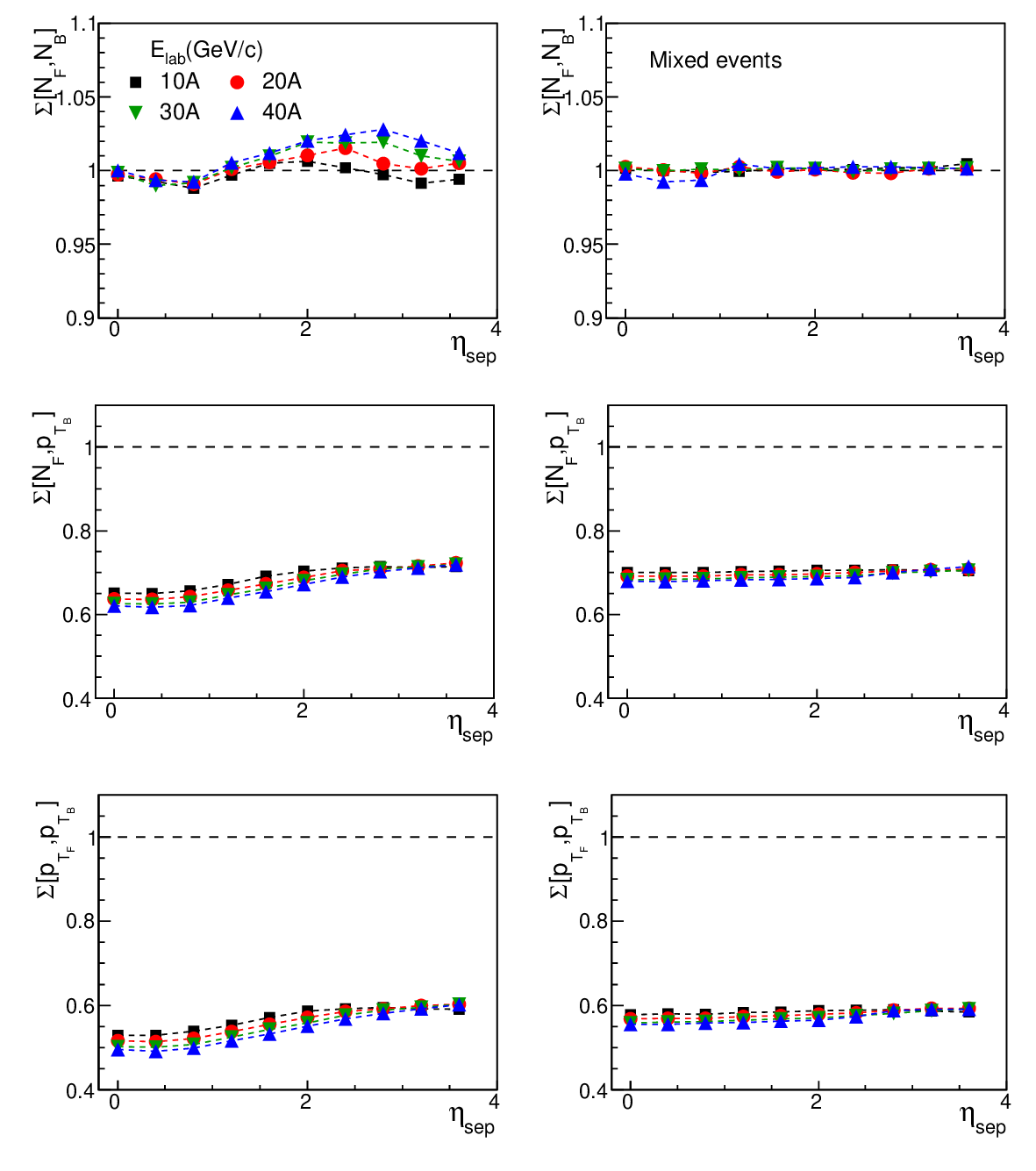}
			\caption{Variations of $\Sigma[N_{\rm F},N_{\rm B}]$, $\Sigma[N_{\rm F},p_{\rm T_B}]$ and $\Sigma[p_{\rm T_F},p_{\rm T_B}]$ with $\eta_{sep}$ for URQMD and corresponding mixed events at FAIR energies.}	
\end{center}							
\end{figure}

\end{document}